\documentclass[12pt,a4paper]{article}

\usepackage[english]{babel}
\usepackage[utf8x]{inputenc}
\usepackage[T1]{fontenc}
\usepackage{blindtext}
\usepackage{enumitem}
\usepackage{float}
\usepackage{commath}
\usepackage{xspace}

\usepackage{subfigure}
\usepackage[toc,page]{appendix}
\usepackage{adjustbox}
\usepackage{amsmath}
\usepackage{graphicx}
\usepackage[colorinlistoftodos]{todonotes}
\usepackage[colorlinks=true, allcolors=blue]{hyperref}
\usepackage{mathtools}
\usepackage{tikz, pgfplots}

\usepackage{authblk}

\makeatletter

\usepackage{algorithm}
\usepackage[noend]{algpseudocode}
\makeatletter
\def\BState{\State\hskip-\ALG@thistlm}
\makeatother

\algdef{SE}[SUBALG]{Indent}{EndIndent}{}{\algorithmicend\ }%
\algtext*{Indent}
\algtext*{EndIndent}

\newlength{\depthofsumsign}
\title{\large{INTRODUCTION TO DATA ASSIMILATION \\ALGORITHMS
FOR PARAMETER ESTIMATION
}}

\date{}
\author{Loc Luong}
\affil{New Mexico Tech}

\begin{document}
\maketitle
\section*{Abstract}
In this study, two classes of methods including statistical and variational data assimilation algorithms will be described. In statistical methods, the model state is updated sequentially based on the previous estimate. Variational methods, on the other hand, seek an estimation in space and time by minimizing a cost function. Both of these methods require estimates of background state which is the prior information of the system and its error covariances. In terms of linear and Gaussian problems, they have the same solution. In the family of Kalman Filter algorithms, the conventional Kalman Filter (KF) and Ensemble Kalman Filter (EnKF) will be implemented. A three-dimension variational method (3D-Var) will be employed to illustrate the variational approaches. A simple case of an ordinary differential equation (ODE) is coupled to highlight the difference between these algorithms. Namely, a mass-spring system governed by a second order differential equation will be examined. We also look at the situation where a periodic external force is applied to the system. Then different data assimilation algorithms will be applied to this system. Results from the experiments will be analyzed to showcase the advantages and disadvantages of each method.

\section{Introduction}

The process of integrating model predictions with actual system measurements to provide an optimal approximation of the current state of the system or predictions of the future of the system is known as data assimilation. In practice, the observations or measurements usually contain noise. It is important to incorporate noisy data in efficient way to provide more accurate estimates of system state. And data assimilation is by far one of the most efficient way to tackle such tasks. This technique combines different types of information which are forecasting data and observation of state to minimize prediction errors. The method initially developed in weather forecasting, and later gain a great deal of attention to other fields of science.\\
Over time, data assimilation became very popular in many fields of science such as geoscience, petroleum engineering, seismology, medicine, control theory, etc. Data assimilation can be classified in two classes: statistical DA and variational DA. In statistical aspect, we seek an optimal solution with minimum variance, whereas in variational DA we seek an optimal solution that minimizes a cost function. Kalman filter is one of the simplest method in data assimilation, and is first developed by R. E. Kalman.

\section{Methodology}
Consider a dynamic system governed by the following equation:
\begin{equation}
x_{k} = Ax_{k-1}+ Bu_{k-1} + w_{k-1}
\end{equation}
where $x_{k}$ is the state vector of the process at time k, A is the state transition model which get the previous state vector $x_{k-1}$ as input, B is the control input model which get the previous control variable $u_{k-1}$ as input, $w_{k-1}$ is the process noise that is assumed to have a multivariate normal distribution with mean 0 and a covariance matrix Q.\\
The measurements or observations $z_{k}$ at time k is modeled by:
\begin{equation}
z_{k} = Hx_{k}+ v_{k}
\end{equation}
where H is the observation model, and is assumed stationary over time, $v_{k}$ is the associated observation error that is assumed to be normally distributed with mean 0 and covariance matrix R.\\
We want to estimate the state of the system in the sense of minimizing the norm of $\norm{z_{k}- Hx_{k}}$ as well as the uncertainty. The Kalman filter helps to model the prior values (background information) of the state and measured values (observations) into the following form:
\begin{equation}
\hat{x}_{k} = \hat{x}'_{k}+ K_{k}(z_{k}- H\hat{x}'_{k})
\end{equation}
where $\hat{x}'_{k}$ is the prior estimate of $x_{k}$, $K_{k}$ is the Kalman gain which will be derived later. The second term $z_{k}- H\hat{x}'_{k}$ is known as the innovation.\\
The covariance of the new update is:
\begin{equation}
P_{k} = cov(\hat{x}_{k})
\end{equation}
\begin{equation}
\begin{split}
P_{k} &= cov((I-K_{k}H)\hat{x}'_{k} + K_{k}z_{k})\\
&= (I-K_{k}H)cov(\hat{x}'_{k})(I-K_{k}H)^{T} + K_{k}cov(z_{k})K_{k}^{T}\\
&= (I-K_{k}H)P'_{k}(I-K_{k}H)^{T} + K_{k}cov(z_{k})K_{k}^{T}\\
&= (I-K_{k}H)P'_{k}(I-K_{k}H)^{T} + K_{k}RK_{k}^{T}\\
\end{split}
\end{equation}
where $P'_{k}$ is the prior estimate of $P_{k}$.\\
Expand the covariance $P_{k}$, we can have:
\begin{equation}
\begin{split}
P_{k} &= P'_{k} - K_{k}HP'_{k} - P'_{k}H^{T}K_{k}^{T} + K_{k}HP'_{k}H^{T}K_{k}^{T} + K_{k}RK_{k}^{T}\\
&= P'_{k} - K_{k}HP'_{k} - P'_{k}H^{T}K_{k}^{T} + K_{k}(HP'_{k}H^{T} + R)K_{k}^{T}
\end{split}
\end{equation}
The sum of diagonal elements of the covariance matrix $P_{k}$ is total variation of the system. Therefore, we want to minimize the trace of the covariance matrix $P_{k}$.\\
Taking the trace operator of both sides, we have:
\begin{equation}
\begin{split}
Tr(P_{k}) &= Tr(P'_{k}) - 2Tr(K_{k}HP'_{k})  + Tr(K_{k}(HP'_{k}H^{T} + R)K_{k}^{T})\\
\end{split}
\end{equation}
Using vector calculus, the derivative with respect to $K_{k}$ can be expressed as: 
\begin{equation}
\begin{split}
\frac{\partial T(P_{k})}{\partial K_{k}} &= -2(HP'_{k})^{T} + 2K_{k}(HP'_{k}H^{T} + R)
\end{split}
\end{equation}
Setting the derivative to be 0, we have:
\begin{equation}
\begin{split}
-2(HP'_{k})^{T} + 2K_{k}(HP'_{k}H^{T} + R) = 0
\end{split}
\end{equation}
This leads to:
\begin{equation}
\begin{split}
K_{k} &= (HP'_{k})^{T}(HP'_{k}H^{T} + R)^{-1}\\
K_{k} &= P'_{k}H^{T}(HP'_{k}H^{T} + R)^{-1}\\
\end{split}
\end{equation}
Then, we can substitute the Kalman gain $K_{k}$ to simplify $P_{k}$ as follows:
\begin{equation}
\begin{split}
P_{k} &= P'_{k} - K_{k}HP'_{k} - P'_{k}H^{T}K_{k}^{T} + K_{k}(HP'_{k}H^{T} + R)K_{k}^{T}\\
&= P'_{k} - P'_{k}H^{T}(HP'_{k}H^{T} + R)^{-1}HP'_{k} - P'_{k}H^{T}(P'_{k}H^{T}(HP'_{k}H^{T} + R)^{-1})^{T}+\\
&+P'_{k}H^{T}(HP'_{k}H^{T} + R)^{-1}(HP'_{k}H^{T} + R)(P'_{k}H^{T}(HP'_{k}H^{T} + R)^{-1})^{T}\\
&= P'_{k} - P'_{k}H^{T}(HP'_{k}H^{T} + R)^{-1}HP'_{k} - P'_{k}H^{T}(P'_{k}H^{T}(HP'_{k}H^{T} + R)^{-1})^{T}+\\
&+P'_{k}H^{T}(P'_{k}H^{T}(HP'_{k}H^{T} + R)^{-1})^{T}\\
&= P'_{k} - P'_{k}H^{T}(HP'_{k}H^{T} + R)^{-1}HP'_{k}\\
&= P'_{k} - K_{k}HP'_{k}\\
&= (I-K_{k}H)P'_{k}
\end{split}
\end{equation}
Finally, the Kalman Filter algorithm can be summarized as follows:

\begin{algorithm}
\caption{Kalman Filter algorithm}\label{euclid}
\hspace*{\algorithmicindent} \textbf{Input:} \text{A, B, H, Q, R, $P_{0}$, $x_{0}$}
\begin{algorithmic}[1]
\State $\textit{For k} \text{ = 1:k}\textit{}$
\Indent
\State $x'_{k} = Ax_{k-1}+Bu_{k}$
\State $P'_{k} = AP_{k-1}A^{T} + Q$
\State $K_{k} = P'_{k}H^{T}(HP'_{k}H^{T} + R)^{-1}$
\State $x_{k} = x'_{k}+K_{k}(z_{k}-Hx'_{k}$.
\State $P_{k} = (I-K_{k}H)P'_{k}$
\EndIndent
\State End For
\end{algorithmic}
\hspace*{\algorithmicindent} \textbf{Output:} \text{$x_{k}$, $P_{k}$}
\end{algorithm}

\newpage
\subsection{Ensemble Kalman Filter}
In the Ensemble Kalman Filter, we use Monte Carlo simulation to generate a collection of the state variable which represent the uncertainty of the system. Firstly, using multivariate normal distribution we can initialize the ensemble for each random vector and perform the estimate update in a way similar to Kalman Filter:
\begin{equation}
x_{k}^{i} = \hat{x}_{k}^{i} + K(y - H\hat{x}_{k}^{i})
\end{equation}
where $\hat{x}_{k}^{i}$ is the forecast state vector, and is estimated as:
\begin{equation}
\hat{x}_{k}^{i} = M(\hat{x}_{k-1}^{i})
\end{equation}
where M is the system operators that can be linear or non-linear.\\
In Kalman Filter, the Kalman gain is:
\begin{equation}
\begin{split}
K_{k} &= P'_{k}H^{T}(HP'_{k}H^{T} + R)^{-1}\\
\end{split}
\end{equation}
We can estimate the Kalman gain using the statistic data from ensemble vectors. Firstly, the forecast covariance $P'_{k}$ can be estimated as:
\begin{equation}
P'_{k} = \frac{1}{m-1}\sum_{i=1}^{m}(\hat{x}_{k-1}^{i} - \bar{x})(\hat{x}_{k-1}^{i} - \bar{x})^{T}
\end{equation}
where $\bar{x}$ is the mean of ensemble vectors:
\begin{equation}
\bar{x} = \frac{1}{m}\sum_{i=1}^{m}\hat{x}_{k-1}^{i}
\end{equation}
Then, the Kalman gain can be calculated similarly to KF using equation 14.\\
The Ensemble Kalman Filter algorithm can be summarized as follows:

\begin{algorithm}[H]
\caption{Ensemble Kalman Filter algorithm}\label{euclid}
\hspace*{\algorithmicindent} \textbf{Input:} \text{Observation model H, Observation covariance matrix R, system operator H}
\begin{algorithmic}[2]
\State \text{Initialize ensemble vectors $\hat{x}_{k}^{i}$, i=1,2..,m.}
\State \textit{For k} \text{ = 1:k}\textit{}
\Indent
\State \text{Perturbation the observation with noise $z_{i,k} = z_{i} + u_{i}$, $u_{i}\sim \mathcal{N}(0,R)$}
\State \text{Compute ensemble mean and covariance}
\State \text{Compute Kalman gain}
\State \text{Update ensemble vectors}
\State \text{Compute ensemble forecast}
\EndIndent
\State End For
\end{algorithmic}
\hspace*{\algorithmicindent} \textbf{Output:} \text{$x_{k}$, $P_{k}$}
\end{algorithm}

\newpage
\subsection{3D-Variational method}
The key concept behind variational methods is to strike a balane between the background (historical state) and the measured values (observations).
In terms of variational approach, we seek for solutions that are not far away from the observation by minimizing a cost function. The cost function is:
\begin{equation}
J(x) = \frac{1}{2}(x-x^{b})^{T}D^{-1}(x-x^{b}) + \frac{1}{2}(y - Hx)^{T}R^{-1}(y- Hx)
\end{equation}
where R and D are the observation and background error covariance matrices, respectively. H is the observation operator.\\
The gradient of J is:
\begin{equation}
\nabla J = D^{-1}(x-x^{b}) - H^{T}R^{-1}(y - Hx)
\end{equation}
Solving equation 18, we can have:
\begin{displaymath}
\begin{split}
D^{-1}(x-x^{b}) &= H^{T}R^{-1}(y - Hx)\\
D^{-1}x +  H^{T}R^{-1}Hx &= H^{T}R^{-1}y + D^{-1}x^{b}\\
x &= (D^{-1} + H^{T}R^{-1}H)^{-1}(H^{T}R^{-1}y + D^{-1}x^{b})\\
x &= (D^{-1} + H^{T}R^{-1}H)^{-1}(D^{-1}x^{b} +H^{T}R^{-1}Hx^{b} + \\
&+H^{T}R^{-1}y - H^{T}R^{-1}Hx^{b})\\
x&= (D^{-1} + H^{T}R^{-1}H)^{-1}((D^{-1} +H^{T}R^{-1}H)x^{b} + \\
&+H^{T}R^{-1}(y - Hx^{b})\\
x&= x^{b} + K(y - Hx^{b})\\
\end{split}
\end{displaymath}
Finally, we have:
\begin{equation}
\begin{split}
x&= x^{b} + K(y - Hx^{b})
\end{split}
\end{equation}
where the gain matrix K is given by:
\begin{displaymath}
\begin{split}
K = (D^{-1} + H^{T}R^{-1}H)^{-1}H^{T}R^{-1}
\end{split}
\end{displaymath}
Equation 19 express the solution of 3D-Variational method. The solution is very similar to the Kalman filter form (Equation 3) where we model the prior values of the state and measured values. The gain matrix K is weighted average of the background (or prior) and the observations of the state.\\
The background covariance matrix D describe the errors covariances of model variables. The matrix D in particular must be symmetric positive definite, and it must be reasonable in expressing observation parameters. In practice, the matrices involved in equation 19 are often in large dimensions, thus the direct calculation is unfeasible. Therefore, to minimize this cost function of 3D-Variation method, gradient descent (GD) is often implemented.
The 3D-Var algorithm is given as:
\begin{algorithm}[H]
\caption{3D-Var algorithm}\label{euclid}
\hspace*{\algorithmicindent} \textbf{Initialization:} \text{$x_{0} = x^{b}$, n =0}
\begin{algorithmic}[1]
\State \text{While $\norm{\nabla J} > \epsilon$}\text{ or $n \leq n_{max}$}
\Indent
\State \text{Compute J}
\State \text{Compute $\nabla J$}
\State \text{Update $x_{j+1}$}
\State \text{j = j+1}
\EndIndent
\State End While
\end{algorithmic}
\hspace*{\algorithmicindent} \textbf{Output:} \text{$x_{k}$}
\end{algorithm}
By using a change of variable, we can reformulate the variational cost function into Tikhonov regularization problem.\\
Let $D = \sigma_{D}^{2}C_{D}$, $R = \sigma_{R}^{2}C_{R}$, and ignore the constant term $\frac{1}{2}$ in the cost function, we have:
\begin{equation}
\begin{split}
J(x) &= (x-x^{b})^{T}D^{-1}(x-x^{b}) + (y - Hx)^{T}R^{-1}(y- Hx)\\
&= \frac{1}{\sigma_{D}^{2}}(x-x^{b})^{T}C_{D}^{-1/2}C_{D}^{-1/2}(x-x^{b}) + \frac{1}{\sigma_{R}^{2}}(y - Hx)^{T}C_{R}^{-1/2}C_{R}^{-1/2}(y- Hx)
\end{split}
\end{equation}
Since $D$ and $R$ are symmetric and positive definite, we obtain:
\begin{equation}
\begin{split}
J(x) &= \frac{1}{\sigma_{D}^{2}}\norm{C_{D}^{-1/2}(x-x^{b})}_{2}^{2} + \frac{1}{\sigma_{R}^{2}}\norm{C_{R}^{-1/2}(y- Hx)}_{2}^{2}\\
&= \frac{1}{\sigma_{D}^{2}}\norm{C_{D}^{-1/2}(x-x^{b})}_{2}^{2}+\frac{1}{\sigma_{R}^{2}}\norm{C_{R}^{-1/2}(y- Hx^{b}) - C_{R}^{-1/2}(Hx- Hx^{b})}_{2}^{2}\\
&= \frac{1}{\sigma_{D}^{2}}\norm{C_{D}^{-1/2}(x-x^{b})}_{2}^{2}+\frac{1}{\sigma_{R}^{2}}\norm{C_{R}^{-1/2}(y- Hx^{b}) - C_{R}^{-1/2}HC_{D}^{1/2}C_{D}^{-1/2}(x- x^{b})}_{2}^{2}\\
\end{split}
\end{equation}
By letting:
\begin{displaymath}
\begin{split}
A &=  R^{-1/2}HD^{1/2}\\
b &= R^{-1/2}(y- Hx^{b})\\
z &= D^{-1/2}(x-x^{b})\\
\mu^{2} &= \frac{\sigma_{R}^{2}}{\sigma_{D}^{2}}
\end{split}
\end{displaymath}
the cost function $J(x)$ became:
\begin{equation}
\begin{split}
J(z) &= \norm{b-Az}_{2}^{2} + \mu^{2}\norm{z}_{2}^{2}
\end{split}
\end{equation}
where $\mu$ can be interpreted as regularization parameter that determined the relative weight between the background state and observations.
It is evident that the classical variational problem can be formulated into the Tikhonov regularization problem as shown above.

\newpage
\section{Results and discussions}
To run experiments, we will consider the a spring-mass system that governed by the following equation:
\begin{equation}
y''(t) + 0.45y'(t) + y(t) = 3cos(2t)
\end{equation}
with initial condition $y(0) =2$, $y'(0) = 0$.\\
By letting:
\begin{equation}
\begin{split}
x_{1}(t) &= y(t)\\
x_{2}(t) &= y'(t)
\end{split}
\end{equation}
we can convert this equation into a system of two first order differential equations. We have:
\begin{equation}
\begin{split}
x_{1}'(t) &= x_2(t)\\
x_{2}'(t) &= -x_{1}(t) - 0.45x_{2}(t) + 3cost(2t)
\end{split}
\end{equation}
This system can be rewritten in:
\begin{equation}
\begin{split}
x'(t) = Ax(t) + Bu(t)
\end{split}
\end{equation}
where
\begin{equation}
A = 
\begin{bmatrix}
0 & 1\\
-1 &-0.45
\end{bmatrix}
\end{equation}
\begin{equation}
B = 
\begin{bmatrix}
1 & 0\\
0 & 1
\end{bmatrix}
\end{equation}
and
\begin{equation}
u(t) = 
\begin{bmatrix}
0 \\
3cost(2t)
\end{bmatrix}
\end{equation}

Using the finite difference approximation method, we can discrete the governing equation and solve for the true state of the system. We assume that the system has process noise with covariance matrix:
\begin{equation}Q = \begin{bmatrix} 0.0005 & 0 \\ 0 &0.0005\end{bmatrix}
\end{equation}, and 
\begin{equation}
H = \begin{bmatrix} 1 & 0 \end{bmatrix}.
\end{equation}
The initial guess is: $x_{0} = \begin{bmatrix} 0\\ 0 \end{bmatrix}$, and $P_{0} = \begin{bmatrix} 0.05 & 0 \\ 0 &0.05\end{bmatrix}$.

\begin{figure}[H]
\centering
\includegraphics[width=1\textwidth]{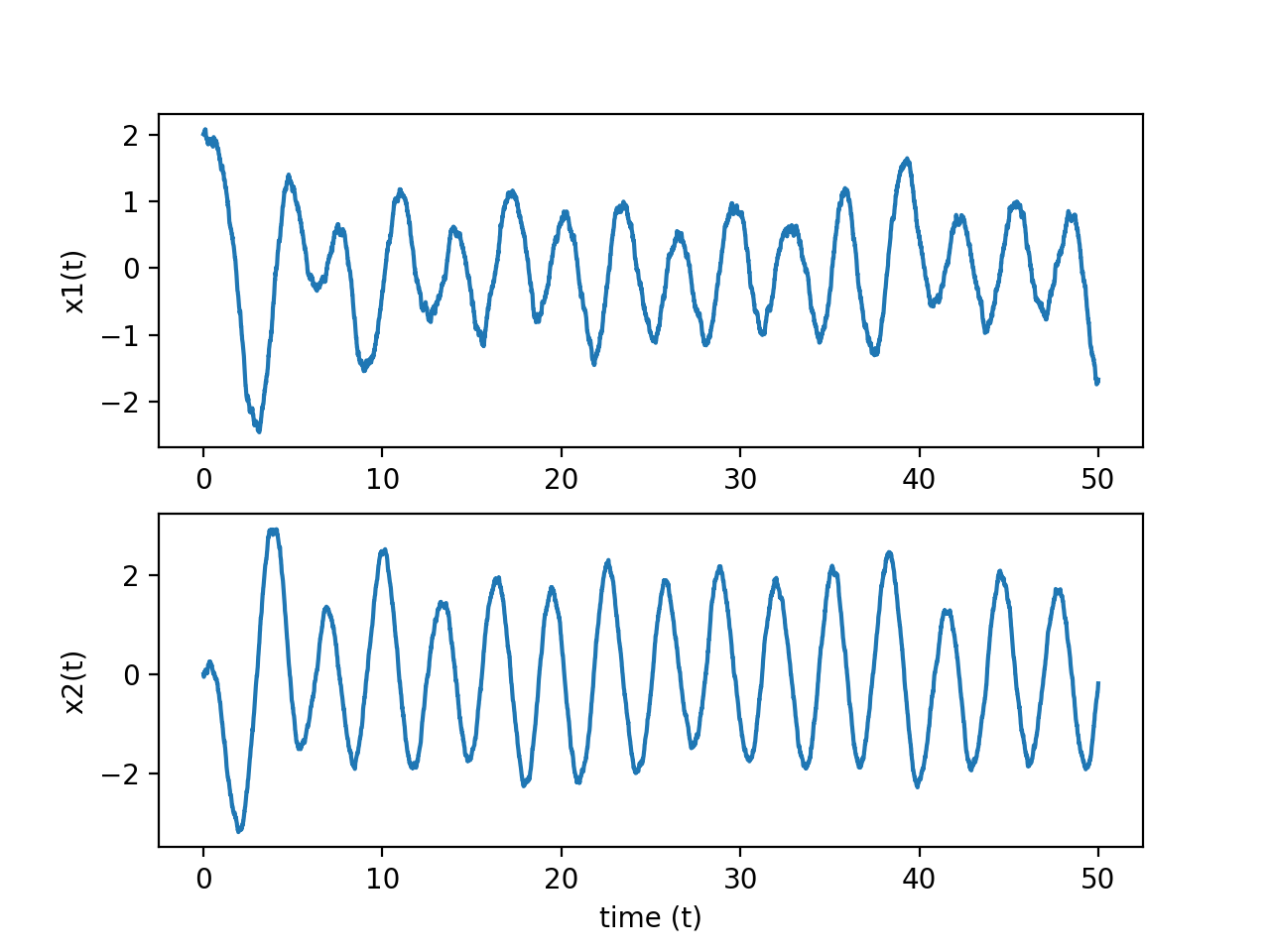}
\caption{The system states x1 and x2}
\end{figure}

\begin{figure}[H]
\centering
\includegraphics[width=1\textwidth]{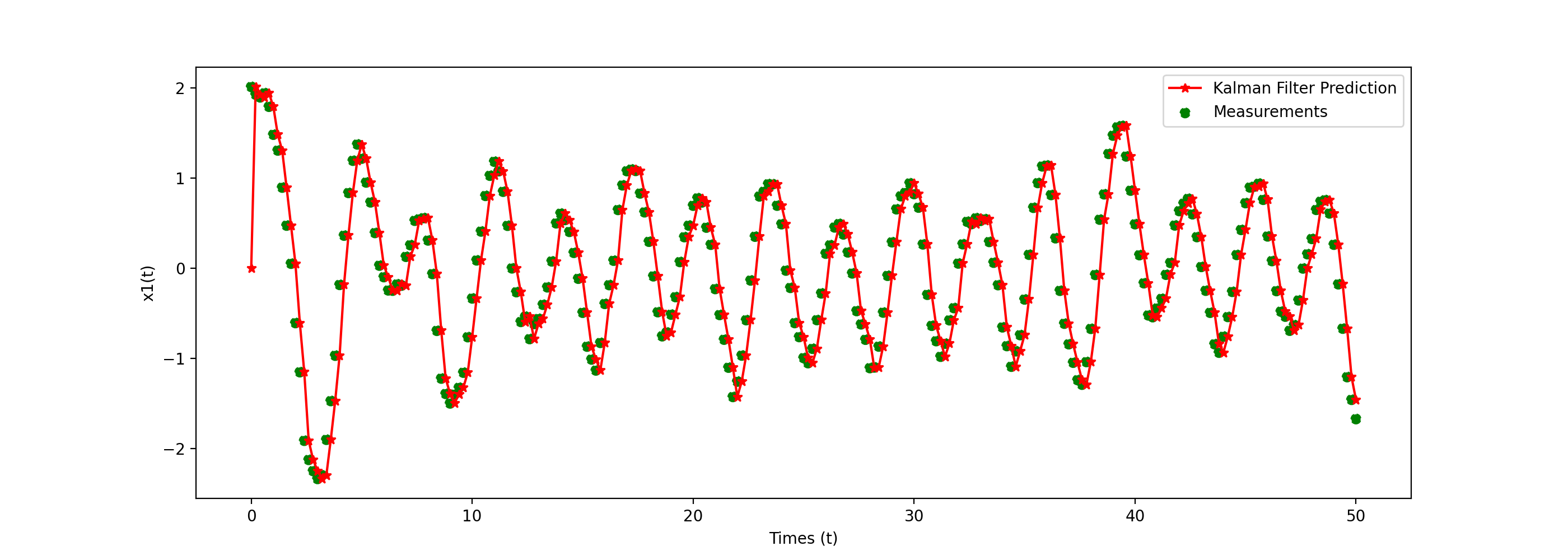}
\caption{Kalman filter prediction}
\end{figure}

\begin{figure}[H]
\centering
\includegraphics[width=1\textwidth]{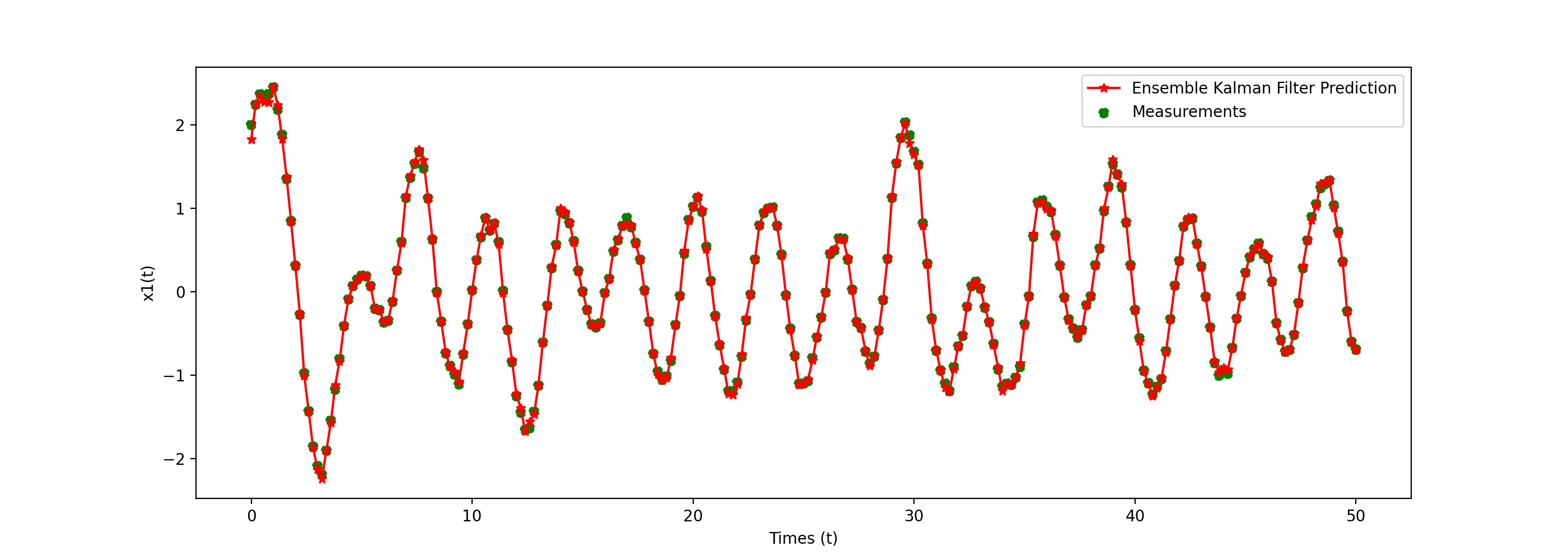}
\caption{Ensemble Kalman filter prediction}
\end{figure}

\begin{figure}[H]
\centering
\includegraphics[width=1\textwidth]{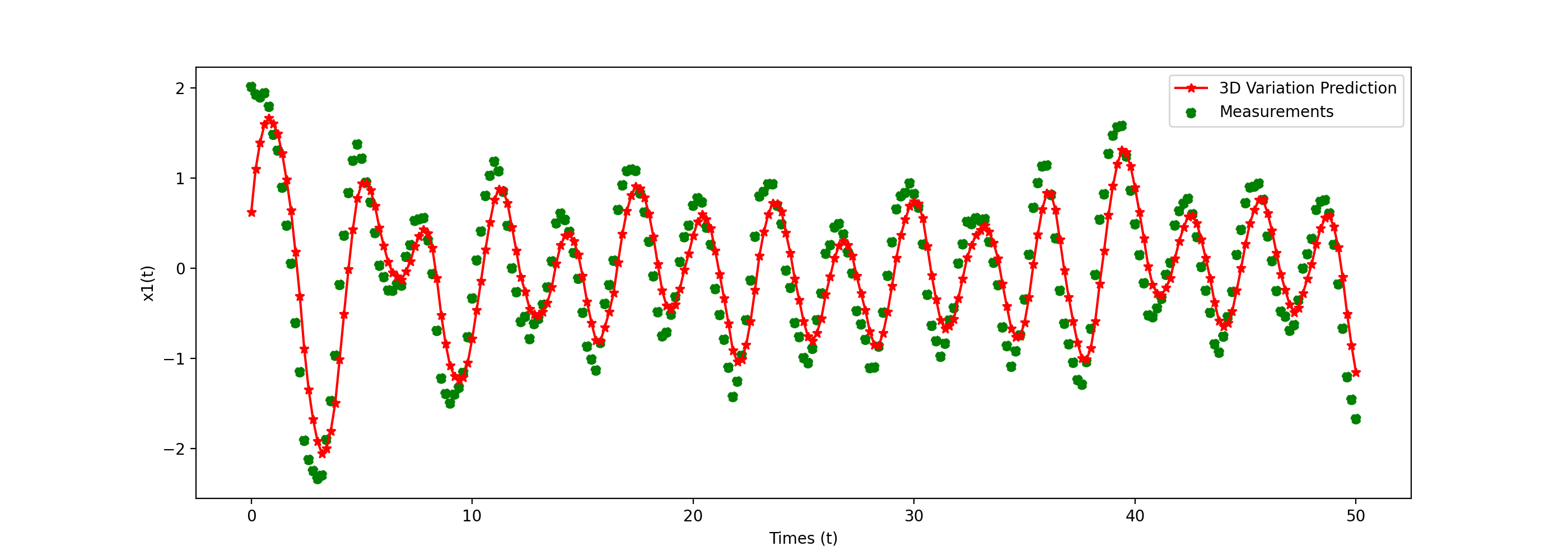}
\caption{3D-Variational prediction}
\end{figure}

Figure 1 depicts the true states of the system. Figure 2 shows the estimate of $x_{1}$ using Kalman Filter (KF) algorithm. Figure 3 illustrates the estimate of $x_{1}$ using Ensemble Kalman Filter (EnKF) algorithm with ensemble size of 100. It is observed that the Ensemble Kalman Filter performed better than Kalman Filter. In both cases, although the initial guesses are far away from observations, after several iterations, the algorithms can be able to quickly get close to the true state of the system.\\
Figure 4 demonstrates the results from 3D-Variational method for predicting $x_{1}$. It can be seen that there are some discrepancies between actual states and predictions from 3D-Var method. This method also takes more iterations to get close to the observations.\\
On the advantage of the Kalman Filter, when new data or observations are available, it is easy to incorporate the previous optimal estimates without recomputing anything. However, the Kalman Filter is not applicable for non-linear system. In contrast, the Ensemble Kalman Filter has been proven in many high-dimensional, non-linear data assimilation problems. By simulating a collection of state vectors this method avoid computing large covariance matrices, and then update the ensemble through time will improve the computational overhead. One of the significant drawbacks of the Ensemble Kalman Filter is that the full error covariance matrix of a high-dimensional system can not be represented with only few ensemble m especially in a rank deficiency problem where m is significant less than n.\\
The 3D-Variational method allows us to solve an optimization problem in one run. A major drawback of this method is the convergence to the global minimum of the cost function to be minimized. Another significant difficulty of 3D-Var method is the need to design an appropriate background covariance matrix D that properly defines the covariances of the model parameters.
\newpage
\section{Conclusion}
In this study, we have implemented different data assimilation algorithms including the Kalman Filter (KF), Ensemble Kalman Filter (EnKF), and 3D-Variational (3D-Var) to solve the spring-mass system governed by a second order differential equations. Results from experiments show that the Ensemble Kalman Filter algorithm has better performance compared to the Kalman Filter algorithm. Although there are some discrepancies in 3D-Var, the method can reasonably predict the states of the system. Results from 3D-Var method are highly depend on the accuracy of the background error covariance matrix D. If D is not define properly, the method may fail to find an optimal solution of the cost function. In further study, we can compare these algorithms with other variants of variational methods as well as filtering methods.


\newpage

\begin{thebibliography}{2}
\bibitem{1} 
Welch, Greg, and Gary Bishop.
\textit{ "An introduction to the Kalman filter." (1995): 127-132.}

\bibitem{2} 
Evensen, Geir. 
\textit{"The ensemble Kalman filter: Theoretical formulation and practical implementation." Ocean dynamics 53, no. 4 (2003): 343-367.}

\bibitem{3} 
Barker, Dale M., Wei Huang, Yong-Run Guo, A. J. Bourgeois, and Q. N. Xiao.
\textit{"A three-dimensional variational data assimilation system for MM5: Implementation and initial results." Monthly Weather Review 132, no. 4 (2004): 897-914.}

\bibitem{4} 
Evensen, Geir.
\textit{Data assimilation: the ensemble Kalman filter. Springer Science \& Business Media, 2009.}

\bibitem{5} 
Asch, Mark, Marc Bocquet, and Maëlle Nodet. 
\textit{Data assimilation: methods, algorithms, and applications. Society for Industrial and Applied Mathematics, 2016.}

\end{thebibliography}
\end{document}